\documentclass[11pt]{article}
\usepackage[textwidth=15.2cm,textheight=22cm]{geometry}
\usepackage{amsmath,amssymb}
\usepackage{latexsym}
\usepackage{multicol}
\usepackage{graphicx}
\usepackage{bm}
\allowdisplaybreaks[1]

\newcommand{\be}{\begin{equation}}
\newcommand{\ee}{\end{equation}}
\newcommand{\ba}{\begin{eqnarray}}
\newcommand{\ea}{\end{eqnarray}}
\newcommand{\bdm}{\begin{displaymath}}
\newcommand{\edm}{\end{displaymath}}

\newcommand\fr[1]{\frac{1}{#1}}

\def\k{\kappa}

\def\ba{\bar A}

\def\beq{\begin{equation}}
\def\eeq{\end{equation}}
\newcommand{\half}{\frac{1}{2}}
\newcommand{\nn}{\nonumber}

\newcommand{\ndt}{\noindent}

\def\bea{\begin{eqnarray}}
\def\eea{\end{eqnarray}}
\def\beas{\begin{eqnarray*}}
\def\eeas{\end{eqnarray*}}
\def\sla{\raise.15ex\hbox{$/$}\kern-.57em}

\def\parm{{\partial}_{-}}

\def\spa#1.#2{\left\langle#1\,#2\right\rangle}
\def\spb#1.#2{\left[#1\,#2\right]}

\begin{document}

\begin{titlepage}
\begin{flushright}    
{\small $\,$}
\end{flushright}
\vskip 1cm
\centerline{\Large{\bf {Cubic interaction vertices in higher spin theories}}}
\vskip 1.5cm
\centerline{Y. S. Akshay and Sudarshan Ananth}
\vskip .5cm
\centerline{\it {Indian Institute of Science Education and Research}}
\centerline{\it {Pune 411021, India}}
\vskip 1.5cm
\centerline{\bf {Abstract}}
\vskip .5cm
Based purely on symmetry considerations, we derive the following result: in  momentum space, the cubic term in the Lagrangian for a field of spin $\lambda$ is equal to the cubic term in Yang-Mills theory, raised to the power $\lambda$. This result is valid for all $\lambda$ for Lagrangians that contain a cubic interaction vertex involving three spin $\lambda$ fields, in four-dimensional flat spacetime. For $\lambda=3$, we present an additional derivation of this result.
\vfill
\end{titlepage}

\section{Introduction}
\ndt A consistent description of the quantum interactions of higher spin fields ($\lambda>2$) is fraught with difficulties, mostly stemming from the higher-derivative structures inherent to such theories. The most comprehensive investigations of higher spin theories~\cite{MV} are at the level of the equations of motion and hence inherently on-shell. We are interested in an off-shell formulation of these theories, with a Lagrangian description~\cite{CF}. Very little is known in this case, even up to the quartic interaction level. The infinite-dimensional gauge symmetries underlying massless higher spin theories could result in surprising ultra-violet properties, making them interesting to study. For a nice summary of various no-go theorems, associated with massless fields in flat spacetime, and ways around them, see~\cite{BBS}. Earlier off-shell  investigations of interacting higher spin theories in flat space include~\cite{BBB,RRM2} and references therein.
\vskip 0.3cm
\ndt {\it {In this paper, we derive the following result: In  momentum space, the coefficient of the cubic interaction vertex for a spin $\lambda$ field is equal to the corresponding Yang-Mills (spin=1) coefficient, raised to the power $\lambda$}}. This result is obtained only for Lagrangians that contain a cubic interaction vertex involving three spin $\lambda$ fields, in four-dimensional flat spacetime. We briefly review the standard light-cone approach to constructing an interacting higher spin theory and then recast the results, introducing spinor helicity products, into a form that makes the result stated above, manifest. Light-cone gauge eliminates the unphysical degrees of freedom but Poincar\'{e} invariance is no longer manifest and will be checked. 
\\
\subsection{Background}

\ndt In this paper, we Fourier transform the higher spin (spin $\lambda$) Lagrangian of~\cite{BBB} and rewrite it in terms of spinor helicity products to prove that the cubic term in the Lagrangian is equal to the cubic term in Yang-Mills theory, raised to the power $\lambda$. One direct consequence of our result: when put on-shell, the coefficient of this cubic term in the higher-spin Lagrangian yields one of the simple scattering amplitude structures suggested in~\cite{BC}.

\subsubsection*{\it {Cubic vertex structure}} We work here with a Lagrangian describing a spin-$\lambda$ field including a cubic interaction vertex involving three spin-$\lambda$ fields~\cite{BBB}. Thus the cubic vertex we consider is only one among a large family of consistent cubic interaction vertices that have been constructed for higher spin theories~\cite{RRM1}. 

\subsubsection*{\it {Spinor helicity and light-cone gauge}} The application of spinor helicity methods to quantum field theory~\cite{LD} has led to significant advances in our understanding of physics. In light-cone gauge, only the physical degrees of freedom ie. helicity states propagate, making it the natural framework to use when applying spinor helicity methods to Lagrangians. 

\subsubsection*{\it {The KLT relations}} Gravity and Yang-Mills theory seem to be decribed be very different Langraigans but there exist surprisingly close perturbative ties between the theories derived starting from the KLT relations~\cite{KLT}. In Yang-Mills theory, we define color-stripped partial amplitudes by
\bea
{\it {A}}^{\rm {tree}}_n=g^{(n-2)}\,{\mbox {Tr}}(\ldots)\,\times\,{\mbox {A}}^{\rm {tree}}_n\ ,
\eea
where ${\it {A}}^{\rm {tree}}_n$ is a tree-level scattering process involving $n$ gluons, $g$ the Yang-Mills coupling and all colour information is contained in the trace. For gravity, we define
\bea
{\it {M}}^{\rm {tree}}_n={{\bigg (}\frac{\kappa}{2}{\biggr )}}^{(n-2)}\,\times\,{\mbox {M}}^{\rm {tree}}_n\ ,
\eea
where ${\it {M}}^{\rm {tree}}_n$ represents a tree-level gravity scattering process and $\k^2=32\pi G_N$ is the coupling in terms of the Newton constant and ${\mbox {M}}^{\rm {tree}}_n$ is the coupling-stripped amplitude. The KLT relations then read
\bea
{\mbox {M}}^{\rm {tree}}_n\propto ({\mbox {A}}^{\rm {tree}}_n)^2\ ,
\eea
suggesting the possibility of a much broader and intriguing relationship of the form
\beas
{\mbox {Gravity}} \sim ({\mbox {Yang-Mills}}) \times ({\mbox {Yang-Mills}})\ .
\eeas
Such a broad relationship however would require far more than simple tree-level connections.

\subsubsection*{\it {MHV Lagrangians}}

This on-shell connection between Yang-Mills theory and gravity does have a nice Lagrangian (off-shell) origin. A suitable canonical field redefinition of the Yang-Mills Lagrangian results in an `amplitude-friendly' MHV Lagrangian wherein the perturbative structures are manifest at the off-shell level of the Lagrangian~\cite{MHV}. An analogous redefinition for gravity~\cite{AT} yields a MHV gravity Lagrangian. The origin of the on-shell KLT relations is then entirely manifest and visible at the level of the respective off-shell (MHV) Lagrangians~\cite{AT}.
\vskip 0.3cm
\ndt {\it {The primary motivation for this work is to ask whether such off-shell realizations of the KLT relations exist even for higher spin theories}}
\vskip 0.5cm
\ndt The aim of the present paper is to begin answering this question. 

\vskip 0.5cm
\ndt We start with a brief review of how the light-cone gauge cubic interaction Lagrangian, for higher spin theories, is derived following~\cite{BBB}. We then Fourier transform and re-write this result in the language of spinor helicity products. After a few simple algebraic manipulations, we will show that the off-shell cubic interaction vertex for a theory of spin $\lambda$ is the corresponding Yang-Mills vertex, raised to the power $\lambda$.

\section{Cubic interaction vertices}
We define light-cone co-ordinates in $(-,+,+,+)$ Minkowski space-time as
\begin{eqnarray}
x^{\pm}=\frac{x^{0}\pm x^{3}}{\sqrt{2}} \;,\qquad
x = \frac{x^{1}+ix^{2}}{\sqrt{2}} \;,\qquad\bar{x}= \frac{x^{1}-ix^{2}}{\sqrt{2}}\ .
\end{eqnarray}
The corresponding derivatives being $\partial_{\pm}\,,\,\,\bar{\partial}$ and $\partial$. In four spacetime dimensions, all massless fields have two physical degrees of freedom $\phi$ and $\bar{\phi}$. The generators of the Poincar\'{e} algebra, in light-cone coordinates, are the momenta
\bea
p^{-}=i\frac{\partial\bar{\partial}}{\partial_{-}}=-p_+ \qquad p^+=-i\partial^{+}=-p_- \qquad p=-i\partial \qquad \bar{p}=-i\bar{\partial}\ ,
\eea
with $\frac{1}{\parm}$ defined using the prescription in~\cite{SM}. The rotation generators
\begin{eqnarray}
j = (x\bar{\partial}-\bar{x}\partial - \lambda) \;,\qquad  j^{+} = i(x^{+}\partial-x\partial_{-})\ , \nn\\
 j^{+-}=i(x^{+}\frac{\partial\bar{\partial}}{\partial_{-}}-x^{-}\partial_{-} ) \;,\qquad j^{-}=i(-x\frac{\partial\bar{\partial}}{\partial_{-}}-x^{-}\partial +\lambda \frac{\partial}{\partial_{-}}) \ ,
\end{eqnarray} 
and their complex conjugates. $\lambda$ is the helicity of the field and $\partial_{+}=\frac{\partial\bar{\partial}}{\partial_{-}}$ using the free equations of motion (which picks up corrections in the interacting theory). 
\\
Any four-vector may be expressed as a bispinor using the Pauli matrices, $p_{a\dot{a}}=p_{\mu}\sigma^{\mu}_{a\dot{a}}$, with $\det(p_{a\dot{a}})$ yielding $-p^{\mu}p_{\mu}$. We also introduce the spinor product
\begin{equation}
<kl>= \sqrt{2}\frac{(kl_{-}-lk_{-})}{\sqrt{k_{-}l_{-}}}\ .
\end{equation}
\ndt The Hamiltonian for the free field theory is
\begin{equation}
H\equiv\int d^{3}x\,\mathcal{H}=-\int d^3x\,\bar\phi\,\partial\bar\partial\,\phi\  ,
\end{equation}
with the last equality valid only for the free theory. We also write
\begin{equation}
\label{hamil}
H\equiv\int d^{3}x\,\mathcal{H}=\int d^{3}x\,\partial_{-}\bar{\phi}\,\delta_{\mathcal{H}}\phi.
\ ,
\end{equation}
thereby introducing the time translation operator
\begin{eqnarray}
\delta_{\mathcal{H}}\phi\equiv\partial_{+}\phi=\lbrace \phi,\mathcal{H}\rbrace
\end{eqnarray}
where $\lbrace ,\rbrace$ denotes the Poisson bracket. When interactions are switched on, this $\delta_{\mathcal H}$ operator picks up corrections, order by order in the coupling constant $\alpha$. At order $\alpha$, based purely on helicity considerations~\footnote{The field $\phi$ has helcity $\lambda$ while the field $\bar\phi$ has helicity $-\lambda$.} , the interacting part of $\delta_{\mathcal{H}}$, based on helicity considerations~\cite{BBB}, has to have the form $\alpha\phi\phi + \alpha \bar{\phi}\phi$ (apart from derivatives and constants).   For the first of these two terms, we start with the following ansatz
\begin{equation}
\label{integers}
\delta^{\alpha}_{\mathcal{H}}\phi=\alpha\,\partial^{+\mu}\left[\bar{\partial}^{a}\partial^{+\rho}\phi\bar{\partial}^{b}\partial^{+\sigma}\phi\right]\  ,
\end{equation}
where $\mu ,\rho ,\sigma ,a,b$ are integers. Three Poincar\'{e} generators pick up $O(\alpha)$ corrections
\begin{eqnarray}
\delta_{j^{+-}}\phi = \delta_{j^{+-}}^{0}\phi -ix^{+}\delta^{\alpha}_{\mathcal{H}}\phi + O(\alpha^{2})\ ,\nn \\
\delta_{j^{-}}\phi = \delta_{j^{-}}^{0}\phi +ix\delta^{\alpha}_{\mathcal{H}}\phi +\delta^{\alpha}_{s}\phi + O(\alpha^{2})\ , \nn \\
\delta_{\bar{j}^{-}}\phi = \delta_{\bar{j}^{-}}^{0}\phi +i\bar{x}\delta^{\alpha}_{\mathcal{H}}\phi +\delta^{\alpha}_{\bar{s}}\phi + O(\alpha^{2})\ ,
\end{eqnarray}
where $\delta_{s}^{\alpha}$ and $\delta_{\bar{s}}^{\alpha}$ represent spin transformations whose forms are not relevant to the results that follow. The requirement of closure of the Poincar\'{e} algebra imposes various conditions on the integers introduced in (\ref {integers}). The commutators
 \begin{eqnarray}
\left[\delta_{j},\delta_{\mathcal{H}}^{\alpha}\right]\phi =0\ , \nn \\
\left[\delta_{j^{+-}},\delta_{\mathcal{H}}^{\alpha}\right]\phi =0\ ,
\end{eqnarray}
yield
\begin{eqnarray}
a+b=\lambda \ , \nn \\
\mu +\rho +\sigma =-1 \ ,
\end{eqnarray} 
while the other commutation relations determine the values of $a$, $b$, $\mu$, $\rho$ and $\sigma$. We thus obtain~\cite{BBB}
\begin{eqnarray}
\delta_{\mathcal{H}}^{\alpha}\phi = \alpha \sum^{\lambda}_{n=0} (-1)^{n}{\lambda \choose n}{(\partial^+)}^{(\lambda -1)}\left[\frac{\bar{\partial}^{(\lambda -n)}}{\partial^{+(\lambda -n)}}\phi \frac{\bar{\partial}^{n}}{\partial^{+n}}\phi\right]\ ,
\end{eqnarray}
for even $\lambda$. For odd-helicity fields, closure of the algebra requires the introduction of an antisymmetric structure constant yielding
\begin{equation}
\delta_{\mathcal{H}}^{\alpha}\phi^{a} = \alpha f^{abc}\sum^{\lambda}_{n=0} (-1)^{n}{\lambda \choose n}{(\partial^+)}^{(\lambda -1)}\left[\frac{\bar{\partial}^{(\lambda -n)}}{\partial^{+(\lambda -n)}}\phi^{b} \frac{\bar{\partial}^{n}}{\partial^{+n}}\phi^{c}\right]\ .
\end{equation}
We repeat the steps above for the $\alpha \bar{\phi}\phi$ term in $\delta_\mathcal{H}$. Using (\ref {hamil}), we may obtain the complete Hamiltonian, to this order. The corresponding Action reads~\cite{BBB}
\be
\label{even}
S=\int d^{4}x \left( \half \bar{\phi}\Box\phi+\alpha \sum^{\lambda}_{n=0} (-1)^{n}{\lambda \choose n}\bar{\phi}{(\partial^+)}^{\lambda}\left[\frac{\bar{\partial}^{(\lambda -n)}}{\partial^{+(\lambda -n)}}\phi \frac{\bar{\partial}^{n}}{\partial^{+n}}\phi\right]\ \right),
\ee
for even $\lambda$ and
\be
\label{odd}
S=\int d^{4}x  \left( \half \bar{\phi}^{a}\Box\phi^{a}+\alpha f^{abc}\sum^{\lambda}_{n=0} (-1)^{n}{\lambda \choose n}\bar{\phi}^{a}{(\partial^+)}^{\lambda}\left[\frac{\bar{\partial}^{(\lambda -n)}}{\partial^{+(\lambda -n)}}\phi^{b} \frac{\bar{\partial}^{n}}{\partial^{+n}}\phi^{c}\right]\right) ,
\ee
for odd $\lambda$.
\\
\ndt Both cubic vertices above, in momentum-space, have the following structure (measure and constants suppressed)
\begin{eqnarray}
&&(\bar{k}_{-}+\bar{l}_{-})^{\lambda} \delta^{4}(p+k+l)
\sum_{n=0}^{\lambda} (-1)^{n}{\lambda \choose n}\left(\frac{\bar{k}}{k_{-}}\right)^{\lambda -n}\left(\frac{\bar{l}}{l_{-}}\right)^{n}\tilde{\bar{\phi}}(p)\tilde{\phi}(k)\tilde{\phi}(l)+c.c. \nn \\
&&=(\bar{k}_{-}+\bar{l}_{-})^{\lambda}\delta^{4}(p+k+l) 
\left(\frac{\bar{k}l_{-}-\bar{l}k_{-}}{k_{-}l_{-}}\right)^{\lambda}\tilde{\bar{\phi}}(p)\tilde{\phi}(k)\tilde{\phi}(l)+c.c.\ .
\end{eqnarray}
The momentum-conserving delta function $\delta^{4}(p+k+l)$ implies that
\begin{eqnarray}
<\bar{l}\bar{p}>=\sqrt{\frac{2}{\bar{p}_{-}\bar{l}_{-}}}(\bar{k}\bar{l}_{-}-\bar{l}\bar{k}_{-}) 
=\frac{\sqrt{\bar{k}_{-}}}{\sqrt{-(\bar{k}_{-}+\bar{l}_{-})}}<\bar{k}\bar{l}>\ , \nn \\
<\bar{p}\bar{k}>=\sqrt{\frac{2}{\bar{p}_{-}\bar{k}_{-}}}(\bar{k}\bar{l}_{-}-\bar{l}\bar{k}_{-})
=\frac{\sqrt{\bar{l}_{-}}}{\sqrt{-(\bar{k}_{-}+\bar{l}_{-})}}<\bar{k}\bar{l}>\ ,
\end{eqnarray}
allowing us to rewrite $(\ref{even})$ and $(\ref{odd})$ as
\be
\label{anseven}
S=\int \frac{d^{4}p}{(2\pi)^{4}}\frac{d^{4}k}{(2\pi)^{4}}\frac{d^{4}l}{(2\pi)^{4}}\hspace{.2cm} (2\pi)^{4}\delta^{4}(p+k+l)\hspace{.2cm}\left(\left[\frac{<\bar{k}\bar{l}>^{3}}{<\bar{l}\bar{p}><\bar{p}\bar{k}>}\right]^{\lambda}\tilde{\bar{\phi}}(p)\tilde{\phi}(k)\tilde{\phi}(l)+c.c.\right)\hspace{0.9cm}
\ee
for even $\lambda$ and
\be
\label{ansodd}
S=\int \frac{d^{4}p}{(2\pi)^{4}}\frac{d^{4}k}{(2\pi)^{4}}\frac{d^{4}l} {(2\pi)^{4}}\hspace{.2cm} (2\pi)^{4}\delta^{4}(p+k+l)\hspace{.2cm} f^{abc}\left(\left[\frac{<\bar{k}\bar{l}>^{3}}{<\bar{l}\bar{p}><\bar{p}\bar{k}>}\right]^{\lambda}\tilde{\bar{\phi}}^{a}(p)\tilde{\phi}^{b}(k)\tilde{\phi}^{c}(l)+c.c.\right)\hspace{0.9cm}
\ee
for odd $\lambda$. We know from the MHV Lagrangian for Yang-Mills theory~\cite{MHV} that the cubic term in the Action looks exactly like the equation above except for the factor in square brackets which instead reads
\bea
{\it L^{\,{\mbox {YM}}}_{\,3}}={{\biggl [}\frac{\spa{\bar k}.{\bar l}^{3}}{\spa{\bar l}.{\bar p}\spa{\bar p}.{\bar k}}{\biggr ]}}\ .
\eea
Thus, the coefficient of the cubic interaction vertex in higher spin Lagrangians is equal to the corresponding coefficient in pure Yang-Mills theory raised to the power $\lambda$, ie.
\bea
\label{result}
{\it L^{\,\lambda}_{\,3}}={{\biggl [}\frac{\spa{\bar k}.{\bar l}^{3}}{\spa{\bar l}.{\bar p}\spa{\bar p}.{\bar k}}{\biggr ]}}^\lambda={\biggl [}{\it L^{\,{\mbox {YM}}}_{\,3}}{\biggr ]}^\lambda\  .
\eea
\vskip 0.5cm

\section{Light-cone Action from the Covariant Action}

\ndt In this section, we relate the light-cone and covariant approaches to higher spin theories for the case $\lambda=3$. We derive (\ref {odd}), for $\lambda=3$, starting from the covariant Action~\cite{s3v}. Our starting point is the free action
\begin{equation}
S_0=\int d^{4}x \left( \frac{1}{2} \phi^{a}_{\mu \nu \rho}\Box \phi^{a\mu \nu \rho}+\frac{3}{2}\partial^{\mu}\phi^{a}_{\mu \sigma \rho}\partial_{\nu}\phi^{a\nu \sigma \rho}+\frac{3}{2}\partial_{\mu}\phi^{a}_{\nu}\partial^{\mu}\phi^{a\nu }
+\frac{3}{4}\partial_{\mu}\phi^{a\mu}\partial_{\nu}\phi^{a\nu }-3\partial_{\mu}\phi^{a}_{\nu}\partial_{\rho}\phi^{a\rho \mu \nu }\right)\ ,
\end{equation}
where $\phi^{a\mu}={\phi^{a\mu \nu}}_\nu$. The corresponding field equations read
\begin{eqnarray}
F_{\mu \nu \rho}^{a}-\frac{3}{2}\eta_{(\mu \nu}F_{\rho)}^{a}=0\ ,
\end{eqnarray}
where $\eta_{\mu\nu}$ is the Minkowski metric and 
\begin{equation}
F_{\mu \nu \rho}^{a}= \Box \phi_{\mu \nu \rho}^{a}-3\partial^{\sigma}\partial_{(\mu}\phi_{\nu \rho )\sigma}^{a}+3\partial_{(\mu}\partial_{\nu}\phi^{a}_{\rho)}\ .
\end{equation}
The Lagrangian and field equations are invariant under a gauge transformation of the form
\begin{eqnarray}
\delta_{\lambda}\phi_{\mu \nu \rho}^{a}=3\partial_{(\mu}\lambda_{\nu \rho)}^{a}\ .
\end{eqnarray}
$\lambda_{\nu \rho}$ is symmetric and traceless implying that nine gauge parameters may be chosen. Light-cone gauge is imposed by setting~\cite{AKHB}
\begin{equation}
\phi^{+\mu \nu a} - \fr{4}\eta^{\mu \nu}{\phi^{+ \sigma}}_\sigma^{a}=0\ . 
\end{equation}
The equations of motion yield
\begin{eqnarray} 
\phi^{++-a}=\phi^{+11a}=\phi^{+22a}=0\ , \nn \\
\phi^{-ija}=\fr{\partial^{+}}\partial_k\phi^{kija}\ ;\qquad \phi^{--ia}=\fr{{\partial^+}^2}\partial_j\partial_k\phi^{ijka}\ , \nn \\ 
\phi^{---a}=\fr{{\partial^+}^3}\partial_i\partial_j\partial_k\phi^{ijka}\ ;\qquad \phi^{11ia}=-\phi^{22ia}\ .
\end{eqnarray}
As a consequence, $\phi^{a\mu}=0$ leaving only two independent components of $\phi^{\mu \nu \rho a}$ written as complex quantities $\phi^{a}=(2)^{-1/2}(\phi^{111a}+i\phi^{112a})$ and $\bar{\phi}^{a}=(2)^{-1/2}(\phi^{111a}-i\phi^{112a})$. Many terms in the interacting covariant action~\cite{s3v} involve $\phi^{\mu a}$ and hence vanish. The non-vanishing terms read
\begin{eqnarray}
S_{\alpha}=\frac{-3}{8}\alpha f^{abc}\int d^{4}x (-2\partial_{\delta}\phi_{\rho \beta \gamma}^{a}\partial^{\epsilon}\phi_{\hspace{.4cm}\sigma}^{b\rho \delta}\phi_{\epsilon}^{c \sigma \beta \gamma}+\phi_{\rho \beta \gamma}^{a}\partial^{\rho} \phi_{\delta \epsilon}^{b}\partial^{\beta}\partial^{\gamma}\phi^{c \delta \epsilon \sigma}-3\partial^{\rho}\phi_{\rho \beta \gamma}^{a}\partial^{\epsilon}\phi^{b\beta \gamma}_{\delta}\partial^{\sigma}\phi^{c\hspace{.2cm}\delta}_{\epsilon \sigma}\nn \\
+3\partial^{\rho}\partial^{\delta}\phi^{a}_{\rho \beta \gamma}\partial^{\sigma}\phi^{b\beta \gamma \epsilon}\phi^{c}_{\epsilon \delta \sigma}+6\partial_{\delta}\phi_{\rho \beta \gamma}^{a}\partial^{\sigma}\phi^{b\rho \beta \epsilon}\partial^{\gamma}\phi_{\epsilon \sigma}^{c\hspace{.2cm}\delta})\hspace{2.95cm}
\end{eqnarray} 
On expansion, the first, third and fourth terms vanish leaving us with
\begin{align}
&S_{0}=\int d^{4}x\hspace{.1cm} 2\bar{\phi}^{a}\Box \phi^{a}+
8 \left[2\left(\bar{\phi}^{a}\partial^{+3}\phi^{b}\frac{\bar{\partial}^{3}}{\partial^{+3}}\phi^{c}+3\bar{\phi}^{a}\frac{\bar{\partial}^{2}}{\partial^{+2}}\phi^{b}\bar{\partial}\partial^{+2}\phi^{c}+3\bar{\phi}^{a}\bar{\partial}^{2}\partial^{+}\phi^{b}\frac{\bar{\partial}}{\partial^{+}}\phi^{c}+\bar{\phi}^{a}\phi^{b}\bar{\partial^{3}}\phi^{c} \right)\right.\\
\nonumber
+&6\left.\left(3\bar{\phi}^{a}\frac{\bar{\partial}^{2}}{\partial^{+}}\phi^{b}\partial^{+}\bar{\partial}\phi^{c}+3\bar{\phi}^{a}\bar{\partial}^{2}\phi^{b}\bar{\partial}\phi^{c}+
\bar{\phi}^{a}\partial^{+2}\phi^{b}\frac{\bar{\partial}^{3}}{\partial^{+2}}\phi^{c}+\bar{\phi}^{a}\partial^{+}\phi^{b}\frac{\bar{\partial}^{3}}{\partial^{+}}\phi^{c}\right)+c.c.\right]
\end{align}
which is (\ref {odd}) for $\lambda =3$.

\begin{center}
* ~ * ~ *
\end{center}

\vskip -0.1cm

\ndt Although the cubic vertices in these theories have such neat structures it remains to be seen if they can be used in a straightforward manner to construct~\cite{lit} higher order interaction vertices~\cite{hov}. Could we expect relations of the form (\ref {result}) to hold at higher orders in the coupling constant? If so, do they define a consistent interacting tree-level S-matrix with the corresponding properties? 

\vskip 0.5cm
\ndt {\it {Acknowledgments}}
\vskip 0.3cm

\ndt We thank Stefano Kovacs, Sunil Mukhi  and Hidehiko Shimada for discussions. This work is supported by the Max Planck Institute for Gravitational Physics through the Max Planck Partner Group in Quantum Field Theory; the Department of Science and Technology, Government of India, through the Ramanujan (SA) and INSPIRE (YSA) fellowships.

\end{document}